\documentstyle[aas2pp4]{article}

\def\,{\thinspace}

\def\kms{km\thinspace s$^{-1}$}

\font\sc=cmr10
\def\HI{{\hbox {H\,{\sc I}}}} 
\def\HII{{\hbox {H\,{\sc II}}}} 
\def\CI{{\hbox {C\,{\sc I}}}} 
\def\CII{{\hbox {C\,{\sc II}}}} 

\received{1997 April}
\accepted{}
\journalid{337}{15 January 1989}
\articleid{11}{14}
\slugcomment{}

\begin{document}
\title{Neutral carbon in the protoplanetary nebulae CRL~618 and
CRL~2688}

\author{K. Young}
\affil{Smithsonian Astrophysical Observatory MS 78, 60 Garden Street,
Cambridge, MA 02138}

\begin{abstract}
The 609~$\rm \mu m$ 
$\rm ^3P_1 \rightarrow {^3P_0}$ line of neutral atomic carbon
has been detected in the protoplanetary nebulae CRL~618 and CRL~2688.
\CI\ appears to be about 7\% as abundant as CO in CRL~2688, and 70\% as
abundant as CO in CRL~618.   \CI\ emission arises primarily from the
slow component of the wind from these objects.   The data presented here
indicate that \CI\ gradually becomes more abundant as the star leaves the
AGB, and \CI\ is significantly enhanced relative to CO before the star
develops a large \HII\ region.
\end{abstract}


\font\sc=cmr9
\def\HI{{\hbox {H\,{\sc I}}}} 
\def\HII{{\hbox {H\,{\sc II}}}} 
\def\CI{{\hbox {C\,{\sc I}}}} 
\def\CII{{\hbox {C\,{\sc II}}}} 

\section{Introduction}

   Shortly after the birth of mm wavelength astronomy, observations of
CO in planetary nebulae (PNe) showed that these objects often have a neutral
envelope as massive or more massive than the optically visible nebula (Mufson
{\it et al.} 1975, Huggins {\it et al.} 1996).   The arrival of large
submillimeter telescopes ten years ago allowed another neutral gas component,
atomic carbon, to be studied via its 609$\mu$ ground-state fine structure line
\CI($1-0$).   Because the fine structure lines of \CI\ arise from transitions
between three rather closely spaced levels, the partition function is
relatively insensitive to temperature throughout a very large range of
plausible temperatures for neutral interstellar material.   This feature
makes is possible to estimate the abundance of \CI\ using only the strength
of the \CI($1-0$) transition.

   Bachiller {\it et al.} (1994) observed \CI($1-0$) in the Ring Nebula, and
found that atomic carbon was several times more abundant than CO in that
object.   Similarly, Young {\it et al.} (1997a) found \CI\ to be more
abundant than CO in the Helix Nebula.   On the other hand, the \CI($1-0$)
transition is barely detectable with current receivers in the prototypical
mass-losing carbon star IRC+10216 (Keene {\it et al.} 1993).  Because AGB
stars such as IRC+10216 are thought to be the progenators of planetary nebulae,
it is natural to ask at what point in the evolution of such an object
does \CI\ become a major constituent in the neutral envelope.   The
transition from the AGB to the PNe stage is thought to take only $\sim 1000$
years (Iben and Renzini 1983), so transition objects, or protoplanetary nebulae
(PPNe), are quite rare.   CRL~2688 (the Egg Nebula) and CRL~618 are two objects
which are believed to be PPNe.

   In the mm regime, CRL~2688 and CRL~618 superficially resemble AGB stars.
Both have thick molecular envelopes, exhibiting strong, optically thick
CO emission.   However in addition to the slow, $\sim 20$ \kms\ wind typical
of an AGB star, these objects also have a fast $\sim 100$ \kms\ wind not
seen in the envelopes of typical AGB stars (Gammie {\it et al.} 1989, Young
{\it et al.} 1992).   In the visible regime, the photospheric emission
from these objects can be seen reflected from dust in the nebula's dense
inner region.   Both of these objects have much hotter photospheres than
would be found on the AGB.
 CRL~2688 appears to be the less-evolved object, because
its spectral type is F5 1a (Crampton {\it et al.} 1975) while that of CRL~618
is a hotter B0 (Westbrook {\it et al.} 1975).  

   In this letter I present \CI($1-0$) spectra of CRL~2688 and CRL~618, and
compare the results with earlier investigations of \CI\ and CO in AGB stars
and planetary nebulae.

\begin{deluxetable}{lrrcccccc}
\tablecaption{$609\mu$ Observations of CRL~618 and CRL~2688 \label{tbl-1}}
\tablewidth{0pt}
\tablehead{
\colhead{Source} & \colhead{$\alpha$} & \colhead{$\delta$} &
\colhead{${\rm V_{LSR}}$} & \colhead{Time$^{\rm \ a}$} &
\colhead{$\sigma^{\rm \ b}$} & \colhead{${\rm F^{\ c}_{LVW}}$} &
\colhead{${\rm F^{\ d}_{HVW}}$} & \colhead{${\rm F^{\ e}_{H_2CO}}$}
\nl
\colhead{} & \colhead{(1950)} & \colhead{(1950)} &
\colhead{(\kms)} & \colhead{(min)} &
\colhead{(K)} & \colhead{} & \colhead{(K \kms)} & \colhead{}
}
\startdata
CRL 618	 & 04:39:34.03 & 36:01:15.9 & -22.3 & 167 & 0.10 & 17.8 & 16  & 8.6\nl
CRL 2688 & 21:00:19.85 & 36:29:44.0 & -33.3 & 179 & 0.05 &  5.1 & --- & 0.8 \nl
\enddata
\tablenotetext{a\ }{Total on--source integration time}
\tablenotetext{b\ }{RMS noise level in blank areas of the spectrum after
removal
of a linear baseline}
\tablenotetext{c\ }{Line temperature integrated over the region where low
velocity CO emission is seen}
\tablenotetext{d\ }{This is the additional flux seen in the ``wings'' of the
CRL 618 spectrum, which may be associated with the high velocity emission
seen in CO spectra. Note
that the total velocity extent of the high velocity CO emission is too large
to fit within the spectrometer's 560 MHz total bandwidth.  The flux reported
here is merely the integrated flux from -160 to 160 \kms, minus the flux
from the low velocity wind region.   This value is extremely uncertain, because
it would be strongly affected by any curvature in the spectrometer's baseline.}
\tablenotetext{e\ }{This is the flux integrated over the velocity range of
the low velocity CO emission, centered on the position where the
H$_2$CO$(7_{17}-6_{16})$ line would fall in the spectrum.}
\end{deluxetable}

\section{Observations and Results}

The observations were made using the 10.4~m telescope of the Caltech
Submillimeter Observatory (CSO) on Mauna Kea, Hawaii, during 1993 November
8--11.   A double sideband 
SIS receiver was used in combination with an AOS spectrometer,
which had a total bandwidth of 560 MHz and a resolution of 1.6 MHz,
corresponding to 1 \kms\ at the frequency of the
$\rm ^3P_1 \rightarrow {^3P_0}$
transition (492.1607 GHz).   The observations of CRL~618 were made in
moderately good weather, during which the single sideband system temperature 
(T$_{\rm SSB}$) was $\sim 5800$ K,
on average.  CRL~2688 was observed during periods of extremely good weather,
during which the (T$_{\rm SSB}$) averaged $\sim 1900$ K.   After every 20
seconds of on--source integration, one of two reference positions, located
$\pm 180\arcsec$ from the source in azimuth, was observed for an equal
time.   In order to minimize any fixed pattern noise in the spectrometer, a
pattern of frequency offsets separated by 5 MHz intervals was applied to
the nominal frequency.  Dithering the frequency in this way had the additional
benefit of removing the sideband ambiguity for any line in the composite
spectrum.  The data were calibrated using the standard hot--sky chopper method,
and corrected for the small (3\%) error introduced by the difference in
sky opacity of the two receiver sidebands.   Saturn was observed to determine
the telescope's main beam efficiency, 46\%.   All temperatures reported here
have been corrected for this efficiency.

\ \CI\ was clearly detected in both objects (see figure~1).   In
addition, there appears to be another spectral line visible in the CRL~618
spectrum, and perhaps very weakly in the CRL~2688 spectrum as well.   
The line's frequency, measured from the CRL~618 profile, 
is $491.973\pm 0.005$ GHz.  The microwave spectroscopy catalogs of Pickett
et al. (1996) and Lovas (1992) were searched for possible identifications
for this line.   The only plausible line within $\pm 0.005$ GHz in either
catalog is H$_2$CO($7_{17}-6_{16}$) (491.968936 GHz), 
106 K above the ground state.   Lower
transisions of H$_2$CO have previously been detected in CRL~618 (Cernicharo
{\it et al.} 1989).   Lines $\sim 100$~K above the ground state can be expected
to appear in CRL~618, which has an excitation temperature of about 90~K
in the region where molecular emission arises (Bujarrabal {\it et al.} 1988).
The detection of the H$_2$CO($7_{17}-6_{16}$) line
 should be regarded as tentative, however, because it lies near the
edge of the spectrum, where the AOS can show spurious features due to the
roll-off of its sensitivity.   Table 1 lists the source
coordinates that were used, the on-source integration time, the velocity
integrated intensity of the \CI($1-0$)
and H$_2$CO($7_{17}-6_{16}$) lines, and the RMS noise
level of the spectra.

    Observations of CO emission lines from CRL~618 and CRL~2688 reveal that
both objects have high velocity outflows in addition to the slow winds
typical of AGB stars (Gammie et al. 1989, Cernicharo {\it et al.} 1989,
Young et al. 1992).   For both
these stars, the velocity extent of the \CI\ emission is nearly identical to
that of the slow wind.  However the ``baseline'' of the CRL~618 spectrum
shows a gentle curvature which could be the result of low-level emission
from the high velocity wind.   The extent of the high velocity wind seen
in CO($3-2$), 380 \kms\ (Gammie et al. 1989), is too large to be entirely
contained within the 560 MHz coverage of the spectrometer used for these
observations, so these spectra cannot be used to accurately calculate the
amount of \CI\ in the high velocity wind material.

\section{Discussion}

   To determine the fraction of carbon in the envelope which is atomic,
it is useful to compare the \CI\ emission with emission from CO molecules,
because CO has the largest abundance of any carbon-bearing molecule.   Such
a comparison is hampered, however, by the fact that low J rotational
transitions of CO have a large optical depth in both CRL~618 and CRL~2688.
One is therefore forced to use spectra from a less abundant species, such
as $^{13}$CO, and then to divide the result by the fractional abundance of
$^{13}$C.

   Bujarrabal {\it et al.} (1994) present $^{13}$CO($2-1$) spectra
of both CRL~618 and CRL~2688 taken at the 30m IRAM telescope.   They
derive an excitation temperature (T$_{\rm ex}$) for several molecular
species, and 
cite a number of references in the literature where estimates of T$_{\rm ex}$
have been made.   While there is considerable

\onecolumn
\begin{figure}
\plotfiddle{figure1.ps_saved}{2in}{-90}{78}{78}{-300}{350}
\caption[]{  The \CI\ spectra of CRL~618 and CRL~2688 are shown.   In each
spectrum the systemic velocity and extent of the Low Velocity Wind, derived
from CO observations (Gammie et al. 1989, Young et al. 1992), are shown.  The
location of the line tentatively identified as H$_2$CO($7_{17}-6_{16}$) is
also shown.
}
\end{figure}
\twocolumn

\noindent scatter amongst the estimates,
I will follow these authors, and adopt T$_{\rm ex} = 20$ K, which falls
comfortably near the middle of the T$_{\rm ex}$ estimates.   For that
temperature, and assuming LTE, the column density of CO is given by

$${\rm N(CO) = {{\it f}(^{12}C)\over {\it f}(^{13}C)} \times 5.6 \times 10^{14}
}$$
$${\rm
\times \int_{-V_o}^{V_o} T_{mb}[^{13}CO(2-1)]\ dV \ cm^{-2}\ \ (1)}$$

\noindent where {\it f}(X) is the fractional abundance of species X, 
${\rm \pm V_o}$ is the velocity extent (in km s$^{-1}$)
over which
emission is seen, and ${\rm T_{mb}}$ is the main beam brightness temperature.
Similarly, for \CI\ in LTE at 20 K:

$${\rm N(\CI) =  1.4 \times 10^{16}
}$$
$${\rm
\times \int_{-V_o}^{V_o} T_{mb}[\CI(1-0)]\ dV \ cm^{-2}\ \ (2)}$$

   These simplistic formulae for column densities assume that the emitting
species is optically thin.   Comparisons of the $^{12}$CO and $^{13}$CO
spectra of CRL~618 and CRL~2688 imply that $^{13}$CO transitions
are not optically thick
in these objects (Yamamura {\it et al.} 1994, Yamamura {\it et al.}
1995), so the error arising from the assumption of low optical depth
in equation 1 is probably small when compared with
the uncertainty in the value of
${\rm {\it f}(^{12}C)/{\it f}(^{13}C)}$.
The rather low ($< 1$ K) peak antenna
temperatures for \CI($1-0$) in these objects are most easily explained if the 
\CI($1-0$) is also optically thin.   However it is possible that the \CI($1-0$)
emission arises from an ensemble of clumps of high optical depth, with a
small combined beam filling factor.   Using equation 2 would then result
in an underestimate of N(\CI).

There is no general agreement on the value of
 ${\rm {{\it f}(^{12}C) / {\it f}(^{13}C)}}$ in carbon stars.   Estimates
range from 8 (Rank {\it et al.} 1974) to 40 (Wannier and Linke 1978).   I'll
adopt the value of 20 obtained by Wannier {\it et al.} (1991), which
was derived using optically thin transitions of carbon-bearing species
in both CRL~618 and CRL~2688.

  The Bujarrabal {\it et al.} (1994) $^{13}$CO($2-1$) spectra are particularly
useful, because their $13''$ beam is quite comparable to the CSO's $15''$
beam at the frequency of \CI($1-0$).   This similarity of beam sizes, combined
with the fact that interferometer maps suggest that both CRL~618 and CRL~2688
are slightly resolved, at least in CO, by a $15''$ beam (Hajian {\it et al.}
1996, Yamamura {\it et al.} 1996), allows the calculation of the 
ratio of \CI\ to CO along the line of sight to these stars.  Entering
the velocity integrated line intensities from  Bujarrabal {\it et al.} and the
integrated intensities from the
\CI($1-0$) spectra presented here, into equations 1 and 2,  I
obtain \CI/CO column density ratios of 0.7 and 0.07 for CRL~618 and CRL~2688
respectively.

    Beichman {\it et al.} (1983) have previously reported the detection of the 
\CI($1-0$) transition in CRL~2688, using for their observations an InSb
bolometer on the KAO, which yielded a $150''$ beam.   They derive
\CI/CO $> 5$, a result that is clearly inconsistent with the value I obtained.
It is possible that \CI\ is geometrically distributed in some way such that
much more emission is detectable with a large beam.   For example, if atomic
carbon were distributed in a large annulus, with a central hole larger than the
CSO's $15''$ beam, then the \CI($1-0$) transition could appear much brighter
with the KAO's large beam.   Because the CO emission arises from a region
comparable in size to the CSO's \CI($1-0$) beam (Yamamura {\it et al.} 1996),
this would mean that \CI($1-0$) emission peaks in a region significantly
displaced from the CO emission.   Such a situation would be unprecedented --
in all stars for which maps of \CI($1-0$)
exist (IRC$+10216$, NGC~6720 and
NGC~7027) the \CI($1-0$) emission peaks at or very near the position of
peak CO emission.   This is true even when \CI\ arises as a product of
the photodissociation of CO and other carbon bearing molecules by interstellar
UV radiation, as in the
case of IRC$+10216$.  Also, the similarity of the shape and width
of the \CI($1-0$)
profile to the CO profiles argues against dramatically different CO
and \CI\ spacial distributions. 
In light of the low signal/noise and very
limited velocity coverage of the Beichman {\it et al.} \CI($1-0$) spectrum,
and considering their nondetection of \CI($1-0$) emission from CRL~618 (which
appears to me to be the brighter object), I suspect that they did not detect
CRL~2688.   However given the strength of the emission reported by Beichman
{\it et al.}, and the tremendous improvement of submillimeter receivers
since 1983, it
would not take much telescope time to confirm their detection using a small
telescope such as 1.2~m Mt. Fuji submillimeter telescope, or
the University of Texas ``focal reducer'' (Plume and Jaffe, 1995), which
converts the CSO into a 1~m telescope.   If the Beichman {\it et al.}
result were confirmed, it would mean the \CI\ distribution in CRL~2688
is quite extraordinary.
\begin{deluxetable}{lcrlcr}
\tablecaption{N(\CI)/N(CO) in Carbon--Rich Stars \label{tbl-2}}
\tablewidth{0pt}
\tablehead{
\colhead{Source} & \colhead{N(\CI)/N(CO)} & \colhead{${\rm T_{eff}}$} &
\colhead{Source} & \colhead{N(\CI)/N(CO)} & \colhead{${\rm T_{eff}}$}
}
\startdata
IRC+10216 & $0.02^{\rm a}$ & $1250^{\rm b}$ &
NGC 7027 & $0.5^{\rm e}$ & 170,000$^{\rm f}$ \nl
CRL 2688  & 0.07 & $6500^{\rm c}$ & 
NGC 6720  & $10^{\rm g}$ & 110,000$^{\rm f}$ \nl
CRL 618  & 0.7  & $30,000^{\rm d}$ &
NGC 7293  & $6^{\rm h}$ & 90,000$^{\rm i}$ \nl
\enddata
\tablenotetext{a\ }{Keene {\it et al.} 1993, N(CO) from their
$^{13}$CO($2-1$) profile and  equation 1}
\tablenotetext{b\ }{Miller, 1970 \hskip 2cm
$^c$\ Spectral type F5 1a  (Crampton {\it et al.} 1975)}
\tablenotetext{d\ }{Spectral type B0 (Westbrook {\it et al.} 1975)\hskip 2cm
$^e$\ Young {\it et al.} 1997b}
\tablenotetext{f\ }{Malkov {\it et al.} 1995\hskip 2cm
$^g$\ Bachiller {\it et al.} 1994}
\tablenotetext{h\ }{ Young {\it et al.} 1997a\hskip 2cm
$^i$\ M\'endez {\it et al.} 1992}
\end{deluxetable}

   Table 2 shows gives the values of N(\CI)/N(CO) for a set of objects
ranging in type from the cool carbon star IRC+10216 to the rather
old PNe NGC 7293 (the Helix Nebula).   These objects are all
carbon--rich, and current theories of stellar evolution suggest that they
are representative of different stages along a single evolutionary path.
It appears that while \CI\ is a very minor constitute of the envelope of
a star on the AGB, it becomes comparable in importance to CO in the
PPNe and early PNe stages, and finally is much more
abundant than CO in a fully matured PNe.

It is interesting to note from table 2 that \break N(\CI)/N(CO)
is higher in CRL~2688 than
in \break IRC+10216.   The calculation of N(\CI)/N(CO) is uncertain enough that
these estimates taken by themselves might not be convincing evidence that
 \CI\ truly
is more abundant relative to CO in CRL~2688 than in IRC+10216.   However
the \CI($1-0$) profile of
IRC+10216 has a dramatically different shape than the CO
profile (Keene {\it et al.} 1993), and \CI\ appears to be confined to a pair
of thin shells in that object.   In contrast, the \CI($1-0$) profile of
CRL~2688 more closely resembles the profile of a CO transition, suggesting
that the optical depth of \CI($1-0$) is higher in this object, and that
\CI\ is not confined to a thin shell.   Since CRL~2688 has not
developed a hot enough photosphere to produce a significant \HII\ region
(Spergel {\it et al.} 1983),
it appears that the photodestruction of molecular carbon reservoirs precedes
the ionization of hydrogen.

    The similarly of the N(\CI)/N(CO) ratios of CRL 618 and NGC~7027 is
somewhat puzzling.  NGC~7027 has a much hotter photosphere, and
a much larger \HII\ region.   In a very young
PNe like NGC~7027, much of the carbon liberated by molecular
photodissociation is quickly ionized and \CII\ becomes extremely prominent
(Liu {\it et al} 1996).
Perhaps the large abundance
of \CI\ seen in mature PNe is arises when the slow dissipation of the
AGB wind allows the UV radiation from the central star to begin destroying
CO and other carbon bearing molecules in clumps throughout the envelope.
 The similarity of the \CI($1-0$)
profile in the Helix to that of CO (Young {\it et al.} 1997a), which is
known to be clumpy (Huggins {\it et al.} 1992), is evidence for this
interpretation.

\acknowledgements

I thank Valentin Bujarrabal, who made several helpful suggestions for
improvements to this letter, and who pointed out some blunders in an
earlier version.   Many thanks are also due to staff of the CSO for their
tireless work in supporting observers at that facillity.
Research at the CSO is supported by National Science Foundation grant
AST93--13929.


\begin{references}{}
\reference{}
Bachiller, R., Huggins, P. J., Cox, P. \& Forveille, T., 1994, A\&A, 281, L93
\reference{}
Beichman, C. A., Keene, Jocelyn, Phillips, T. G., Huggins, P. J., Wooten,
H. A., Masson, C. \& Frerking, M. A., 1983, ApJ, 273, 633.
\reference{}
Bujarrabal, V., G\'omez-Gonz\'alez, J., Bachiller, R. \& Martin-Pintado, J.,
1988, A\&A, 204, 242
\reference{}
Bujarrabal, V., Fuente, A. \& Omont, A., 1994, A\&A, 285, 247
\reference{}
Cernicharo, J., Gu\'elin, M., Martin-Pintado, J., Pe\~nalver, J. \&
Mauersberger, R., 1989, A\&A, 222, L1
\reference{}
Crampton, D., Cowley, A. P. \& Humphreys, R. M., 1975, ApJ, 198, L135
\reference{}
Gammie, C. F., Knapp, G. R., Young, K., Phillips, T. G. \& Falgarone, E.,
	1989, ApJ, 345, L87
\reference{}
Hajian, Arsen R., Phillips, J. A. \& Terzian, Yervant, 1996, ApJ 467, 341
\reference{}
Huggins, P. J., Bachiller, R., Cox \& P. Forveille, T., 1992, ApJ, 401, L43
\reference{}
Huggins, P. J., Bachiller, R., Cox, P., \& Forveille, T., 1996, A\&A, 315,
284
\reference{}
Iben, I. \& Renzini, A., 1983, {\it Ann. Rev. Astron. Astrophys}, 21, 271.
\reference{}
Keene, J., Young, K., Phillips, T.J., Buttgenbach, T.H. \&
	Carlstrom, J.E. 1993, ApJ, 415, L131
\reference{}
Liu, X. -W., Barlow, M. J., Nguyen-Q-Rieu, Truong-Bach, Cox, P., Pequignot, D.,
Clegg, P. E., Swinyard, B. M., Griffin, M. J., Baluteau, J. P., Lim, T.,
Skinner, C. J., Smith, H. A., Ade, P. A. R., Furniss, I., Towlson, W. A.,
Unger, S. J., King K. J., Davis, G. R., Cohen M., Emery R. J., Fischer, J.,
Glencross, W. M., Caux E., Greenhouse, M. A., Gry C., Joubert, M.,
Lorenzetti, D., Nisini, B., Omont, A., Orfei, R., Saraceno, P., Serra, G.,
Walker, H. J., Armand, C., Burgdorf, M., Di Giorgio, A., Molinari, S.,
Price, M., Texier, D., Sidher, S. \& Trams, N., 1996, A\&A, 315, 257L 
\reference{}
Lovas, Frank J., 1992 J. Phys. Chem. Ref. Data, 21, 181
\reference{}
Malkov, Y. F., Golovatyj, V. V. \& Rokach, O. V., 1995, Ap\&SS, 232, 99
\reference{}
M\'endez, R. H., Kudritzki, R. P. \& Herrero, A., 1992, A\&A, 260, 329
\reference{}
Miller, Joseph S., 1970, ApJ, 161, L95
\reference{}
Mufson, S. L., Lyon, J. \& Marionni, P. A., 1975, ApJ, 201, L85.
\reference{}
Pickett, H. M., Poynter, R. L., Cohen, E. A., Delitsky, M. L.,
	Pearson, J. C. \& M\"uller, H. S. P., 1996, JPL publication 80--23,
	Rev. 4.
\reference{}
Plume, R \& Jaffe, D. T., 1995, PASP, 107, 488
\reference{}
Rank, D. M., Geballe, T. R. \& Wollman, E. R., 1974, ApJ, 187, L111
\reference{}
Spergel, D. N., Giuliani, J. K. \& Knapp, G. R., 1983, ApJ. 275, 330
\reference{}
Wannier, P. G. \& Linke, R. A., 1978. ApJ, 225, 130
\reference{}
Wannier, P. G., Andersson, B--G, Olofsson, H., Ukita, N. \& Young, K.,
1991, ApJ, 380, 593
\reference{}
Westbrook, W. E., Becklin, E. E., Merrill, K. M., Neugebauer, G., Schmidt,
M., Willner, S. P. \& Wynn--Williams, C. G., 1975, ApJ 202, 407
\reference{}
Yamamura, Issei, Shibata, Katsunori M., Kasuga, Takashi \& Deguchi, Shuji,
1994, ApJ, 427, 406
\reference{}
Yamamura, Issei, Onaka, Takashi, Kamijo, Fumio, Deguchi, Shuji \&
Ukita, Nobuharu, 1995, ApJ, 439, L13
\reference{}
Yamamura, Issei, Onaka, Takashi, Kamijo, Fumio, Deguchi, Shuji \&
Ukita, Nobuharu, 1996, ApJ, 465, 926
\reference{}
Young, K., Serabyn, G., Phillips, T. G., Knapp, G. R., G\"usten, R. \&
	Schulz, A., 1992, ApJ, 385, 265
\reference{}
Young, K., Cox, P., Huggins, P. J., Forveille, T. \& Bachiller, R., 1997,
ApJ, {\it in press} (1997a)
\reference{}
Young, K., Keene, J., Phillips, T.G., Betz, A.L. \& Boreiko, R.T. 
	1997, ApJ, in preparation (1997b)
\end{references}
\end{document}